\documentstyle[12pt,epsf]{article}

\def\mypagenumber{1}
\def\mydate{August 27, 1996}
\def\myend{\end{document}}

\normalsize

\newcounter{sxn}

\newcounter{axn}

\date{}
\newdimen\mybaselineskip
\newcommand{\beeq}{\begin{equation}}
\newcommand{\eneq}{\end{equation}}
\newcommand{\beqn}{\begin{eqnarray}}
\newcommand{\eeqn}{\end{eqnarray}}

\def\mybig{\displaystyle \strut }

\def\dd{\partial}
\def\la{\raise.16ex\hbox{$\langle$} \, }
\def\ra{\, \raise.16ex\hbox{$\rangle$} }
\def\ran{\raise.16ex\hbox{$\rangle$} }
\def\go{\rightarrow}

\def\next{{~~~,~~~}}
\def\onehalf{ \hbox{${1\over 2}$} }
\def\onethird{ \hbox{${1\over 3}$} }
\def\psibar{ \psi \kern-.65em\raise.6em\hbox{$-$}\lower.6em\hbox{} }
\def\mbar{ m \kern-.78em\raise.4em\hbox{$-$}\lower.4em\hbox{} }

\def\Bbar{ B \kern-.73em\raise.6em\hbox{$-$}\hbox{} }

\def\ep{\epsilon}

\def\wil{ \Theta_{\rm W} }
\def\Pw{ P_{\rm W} }

\def\tot{{\rm tot}}

\def\mass{{\rm mass}}

\def\vac{{\rm vac}}

\def\for{\hbox{for }}

\def\myfrac#1#2{{\mybig #1\over \mybig #2}}

\def\boxit#1{$\vcenter{\hrule\hbox{\vrule\kern3pt
     \vbox{\kern3pt\hbox{#1}\kern3pt}\kern3pt\vrule}\hrule}$}
\def\bigbox#1{$\vcenter{\hrule\hbox{\vrule\kern5pt
     \vbox{\kern5pt\hbox{#1}\kern5pt}\kern5pt\vrule}\hrule}$}

\begin{document}

\bibliographystyle{unsrt}

\footskip 1.0cm
\thispagestyle{empty}
\setcounter{page}{\mypagenumber}

{\baselineskip=10pt \parindent=0pt \small
\mydate 
\hfill \hbox{\vtop{\hsize=3.8cm \small 
      UMN-TH-1504/96\\ NUC-MINN-96/13-T \\}}
}

\vspace{35mm}

{\baselineskip=40pt
\centerline {\LARGE\bf Anomalous Behavior}
\centerline {\LARGE\bf in the Massive Schwinger Model}
}

\vspace*{30mm}
\centerline {\large  
Yutaka Hosotani\footnote{e-mail: yutaka@mnhepw.hep.umn.edu} and
Ram\'on Rodr{\'\i}guez\footnote{e-mail: rodriguez@mnhepo.hep.umn.edu}}

\vspace*{10mm}
\centerline {\small\it School of Physics and Astronomy, University
       of  Minnesota}
\centerline {\small\it Minneapolis, Minnesota 55455, U.S.A.}
\bigskip
\vspace*{20mm}
\normalsize

\begin{abstract}

\baselineskip=17pt
We evaluate the chiral condensate and  Polyakov loop
in two-dimensional QED with a fermion of an arbitrary mass ($m$).  We find
discontinuous $m$ dependence in the chiral condensate and anomalous
temperature  dependence in Polyakov loops when the vacuum angle 
$\theta$$\sim$$\pi$ and $m$=O($e$).  These nonperturbative phenomena are due to
the bifurcation process in the solutions to the vacuum eigenvalue equation. 
\end{abstract}

\vspace*{5mm}

\newpage
\baselineskip=22pt


The Schwinger model, QED in two dimensions 
has been a preferred theoretical laboratory for the study of physical 
phenomena such as chiral symmetry, gauge symmetry, anomalies, and 
confinement. \cite{Schwinger}-\cite{Nielsen}  In a nontrivial topology 
it allows us to inquire about 
finite volume and temperature effects while keeping the computations 
infrared safe. \cite{Wolf}-\cite{RHHI}  Results at finite temperature ($T$) can
be obtained by Wick  rotating the solution on a circle $S^1$ of circumference
$L$ and replacing
$L$ by $T^{-1}$.

The theory is exactly solvable with massless fermions,  but not with 
 massive fermions.   The effect of a small
fermion mass ($m/e \ll 1$) in the one flavor case is minor other than 
necessiating the $\theta $ vacuum.\cite{CJS,Coleman1,HH} The opposite limit 
of weak coupling, or  heavy fermions, has been analized by
Coleman. \cite{Coleman2,Coleman1}  In this work we investigate
physical  quantities such
as chiral condensate and Polyakov loop  with no restriction on values of the
parameters of the system. The effect of nonvanishing fermion masses has been
investigated in lattice gauge theory and light cone quantization methods
as well. \cite{Irving}-\cite{Harada}

The Lagrangian of the system is given by 
\beqn
&&{\cal L} = - \hbox{$1\over 4$} \, F_{\mu\nu} F^{\mu\nu} + 
\psibar \gamma^\mu (i \dd_\mu - e A_\mu)  \psi  
- m ( M + M^\dagger )\cr
&&M = \psibar \onehalf(1-\gamma^5)\psi , 
\label{Lagrangian}
\eeqn
where $\gamma^\mu = (\sigma_1, i\sigma_2)$ and 
$\psi_a^T=(\psi^a_+,\psi^a_-).$
We study the model on a circle of circumference $L$ and boundary conditions 
\beqn
A_\mu(t,x+L) &=& A_\mu(t,x) \cr
\psi_a(t,x+L) &=& - \psi_a(t,x) ~.
\label{BC}
\eeqn  
The only physical degree
of freedom associated with gauge fields is the Wilson line phase 
$\wil(t)$: \cite{Manton,HH,Iso}
\beeq
e^{i\wil(t)} = \exp \bigg\{ ie\int_0^L dx \, A_1(t,x) \bigg\} ~~.
\label{Wilson1}
\eneq

In the Matsubara formalism of finite temperature field theory 
boson and fermion fields are  
periodic and anti-periodic in imaginary time ($\tau$),  respectively. 
Mathematically,  the model at finite temperature $T=\beta^{-1}$ is
obtained from the model defined on a circle by Wick rotation and replacement
$L \go \beta$, $it \go x$ and $x \leftrightarrow \tau$.  
The Polyakov loop of a charge $q$ in the finite temperature 
theory corresponds to the Wilson line phase:
\beeq
P_q(x) = \exp \bigg\{ iq\int_0^\beta d\tau \, A_0(\tau, x) \bigg\}
\Longleftrightarrow \exp \bigg\{ i\, {q\over e} \, \wil(t) \bigg\}~~.
\label{PWcorrespondence}
\eneq

We bosonize the fermion in the Coulomb gauge 
in the interaction picture defined by a massless fermion:\cite{HH,HHI}  
\beqn
&&\psi_\pm (t,x) = {1\over \sqrt{L}} \, C_\pm \,
 e^{\pm i \{ q_\pm + 2\pi p_\pm (t \pm x)/L \} }
  :\, e^{\pm i\sqrt{4\pi} \phi_\pm (t,x) } \, : ~ ,\cr
\noalign{\kern 10pt}
&&C_+ =1, \next C_- = \exp \{ i\pi ( p_+ - p_- ) \} ~ ,\cr 
\noalign{\kern 10pt}
&&\phi_\pm (t,x) = \sum_{n=1}^\infty (4\pi n)^{-1/2} \,
\big\{ c_{\pm,n} \, e^{- 2\pi in(t \pm x)/L} + {\rm h.c.} \big\} ~ ,\cr
\noalign{\kern 10pt}
&&e^{2\pi i p_\pm} ~ | \, {\rm phys} \ra = |\, {\rm phys} \ra ~ ,
\label{bosonize}
\eeqn
where $[q_\pm, p_\pm] = i,$ and 
$[c_{\pm,n}, c^{\dagger}_{\pm,m}] =  \delta_{nm}.$
The $:~:$ in  (\ref{bosonize})  indicates normal ordering with respect to
$(c_n^{},c_n^\dagger)$. In physical states $p_\pm$ takes an integer
eigenvalue.

The Hamiltonian in the Schr\"odinger picture becomes
\beqn
&&H_\tot= H_0 + H_\phi  + H_\mass - {\pi\over 6L}\cr
\noalign{\kern 8pt} 
&&H_0~  =  {e^2 L\over 2} \Pw^2 
 + {1\over 2\pi L} \Big\{ \wil + 2\pi p \Big\}^2 \cr
\noalign{\kern 8pt} 
&&H_\phi = \int_0^L dx \, {1\over 2} :\,  \bigg[ ~ \Pi^2 + (\phi ')^2 
+ {e^2 \over\pi }{\phi }^2 ~ \bigg] :\,  \cr
\noalign{\kern 8pt}
&&H_\mass = \int_0^L dx \, m \, (M + M^\dagger ) ~~~.
    \label{Hamiltonian}
\eeqn
The conjugate pairs are
$\{p,q\}=\big\{ \onehalf (p_+ + p_-),  q_+ + q_-  \big\}$, 
$\{\tilde p, \tilde q\} =\big\{ p_+ - p_-,\onehalf (q_+ -
q_-) \big\}$, $\{\Pw, \wil\}$, and $\{ \Pi, \phi=\phi_++\phi_- \}$.
Note that $(\phi, \Pi)$ fields are subject to conditions $\int_0^L dx \,
\phi(x) = 0 = \int_0^L dx \, \Pi(x)$.
The mass operator is given by
\beeq
M =
  - C^{\dagger}_- C_+\cdot e^{ - 2\pi i \tilde p x/L}
\, e^{i q}  \cdot 
 L^{-1}  N_0[e^{i \sqrt{4\pi} \phi}] 
\label{massOperator}
\eneq
where $N_\mu [\cdots]$ indicates that the operator inside $[ ~~]$ is
normal-ordered with respect to a mass $\mu$.

As $[\tilde p, H_\tot] =0$, we may restrict ourselves to states with
$\tilde p=0$.  $p$ takes integer eigenvalues in this subspace.
The Hamiltonian (\ref{Hamiltonian}) posseses a residual gauge symmetry  
$\wil \go \wil +2\pi $ and $p \go p -1$ generated by $U$ defined as\cite{HH}
\beeq
U = \exp \Big( 2\pi i\Pw + iq \Big) ~~~,~~~
[U, H_\tot] = 0 ~~.
\label{GaugeInv1}
\eneq
 The ground state is the $\theta $ vacumm: \cite{Lowenstein}-\cite{Coleman1}
\beeq
U \, |\Phi_\vac (\theta )\ra =  
e^{i\theta } \, |\Phi_\vac (\theta ) \ra ~~ .
\label{vacuum1}
\eneq

To determine the vacumm wave function we must solve the eigenvalue 
equation  
\beeq
(H_0+H_\mass) |\Phi_\vac(\theta) \ra = E |\Phi_\vac(\theta) \ra ~~. 
\label{Schro1}
\eneq
The fermion mass term $H_\mass$ changes the mass of the boson field
$\phi$ from $\mu=e/\sqrt{\pi}$ to $\mu_1$.  The vacuum is defined with respect
to the physical boson mass $\mu_1$.  
Making use of \cite{HH,Sachs}
\beqn
&&N_0[e^{i\sqrt{4\pi}\phi}] = B(\mu L) \, N_{\mu} 
[ e^{i\sqrt{4\pi }\phi } ]  \cr
\noalign{\kern 10pt}
&&B(z) =
{z\over 4\pi} \exp \Bigg\{ \gamma + {\pi\over z}
 - 2 \int_1^\infty  {du \over (e^{uz} - 1)\sqrt{u^2-1}}  \Bigg\}~, 
\label{massOperator2}
\eeqn
the mass operator in (\ref{massOperator}) is  accordingly written as   
\beeq
M =  - C^{\dagger}_- C_+\cdot e^{ - 2\pi i \tilde p x/L}
\, e^{i q}  \cdot 
 L^{-1} B(\mu_1 L)  N_{\mu_1} [e^{i \sqrt{4\pi} \phi}] ~.
\label{massOperator3}
\eneq

With this understanding we write the
vacuum wave function, taking (\ref{vacuum1}) into account, as
\beqn
&&|\Phi_\vac (\theta )\ra = {1\over \sqrt{2\pi}} \sum_n
\int dp_W ~ |p_W, n\ra \, e^{-in\theta+ 2\pi inp_W} ~ f(p_W) \cr
&&\int dp_W ~ |f(p_W) |^2 = 1 ~~~.
\label{vacuum2}
\eeqn
Here $|p_W, n\ra$ is an eigenstate of $P_W$ and $p$.   Since
$\la p_W', n'| e^{\pm iq} |p_W,n\ra
= \delta(p_W'-p_W) ~ \delta_{n',n\pm 1}$, 
the vacuum eigenvalue equation (\ref{Schro1}) is reduced to a
Schr\"odinger equation
\beeq
\bigg\{ -{d^2\over dp_W^2} + V(p_W) \bigg\} ~f(p_W) = \ep ~f(p_W)
\label{Schro2}
\eneq
where
\beqn
&&V(p_W) = \omega^2 p_W^2  -\kappa \cos(\theta - 2\pi p_W) 
\cr
\noalign{\kern 6pt}
&&\omega= \pi\mu L ~~~,~~~
\kappa = 4\pi mL B(\mu_1 L) ~~~,~~~
\mu^2 = {e^2\over \pi} 
\label{Schro3}
\eeqn
and $\ep = 2\pi LE_\vac + \onethird \pi^2$.

To determine the boson mass $\mu_1$, we expand $H_\mass$ in
(\ref{Hamiltonian}) in power series in $\phi$.  In the vacuum
\beqn
H_\mass &\go&  \int_0^L dx ~ {2 \pi m \over  L} ~B(\mu_1 L) 
\la e^{iq} + e^{-iq} \ra_\vac ~ \phi^2 \cr
\noalign{\kern 8pt}
&=& \int_0^L dx ~ {4\pi m B(\mu_1 L)\over L} ~
\la \cos(\theta - 2\pi p_W)\ra_f ~ \phi^2
\label{bosonmass1}
\eeqn
where
$\la F(p_W) \ra_f = \int dp_W ~ F(p_W) \, |f(p_W) |^2$.  Hence
\beeq
\mu_1^2 = \mu^2 + {8\pi m B(\mu_1 L)\over L} ~
\la \cos(\theta - 2\pi p_W)\ra_f ~~~.
\label{bosonmass2}
\eneq
(\ref{Schro2}) and (\ref{bosonmass2}) must be solved simultaneously.
We have a Schr\"odinger problem in which the potential needs to be
determined selfconsistently. \cite{HHI}
Wave functions $f(p_W)$ for typical values for $T/\mu$, $m/\mu$, and $\theta$
are displayed in fig.\ 1.  

\begin{figure}[t]
\hskip 3cm
\epsfxsize= 8.5cm
\epsffile[96 252 384 668]{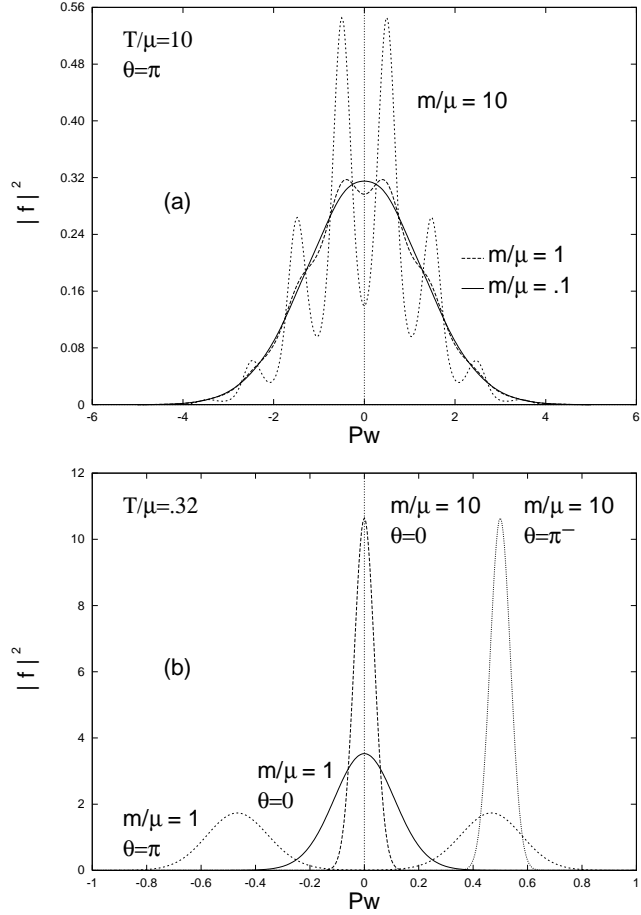}
\vskip 0.cm 
\caption{Wave functions $|f(p_W)|^2$.
(a) $m/\mu$= 10, 1, and 0.1 corresponds to 
$\kappa = 3.61\cdot 10^{-1}, ~ 3.19\cdot 10^{-2}$, and $3.18\cdot
10^{-3}$. 
(b) $\theta=\pi^-$ indicates that the $\theta$ value is less than, but
 is very close to $\pi$.
At $m/\mu$=1,  $\theta$=0 and $\pi$ corresponds to $\kappa$= 2.24 and 2.19. 
At $m/\mu$=10,  $\kappa$=165 for both $\theta$=0 and $\pi$.}
\label{fig.1}
\vskip 0.3cm
\end{figure}

The chiral condensate is given by\footnote{~ We have defined the mass operator
$M$ by (\ref{massOperator}), independent of a mass $m$. 
Consequently $\la \psibar\psi\ra$$\not$=0 
even in the limit $e$$\go$0.  It may be appropriate
to define a physical $M$  by 
$M^{\rm phys}$=$M$$-$$ \la M \ra_{m,e=0}$ so that 
$\la \psibar\psi \ra^{\rm phys}_m$ vanishes in the free theory.
The composite operator $M^{\rm phys}$ thus defined depends on $m$.}
\beqn
\la \psibar\psi \ra_\theta &=& \la M+M^\dagger \ra_\theta \cr
&=& - 2 L^{-1} B(\mu_1 L) \, \la \cos(\theta-2\pi p_W)\ra_f ~~.
\label{condensate1}
\eeqn
Combining (\ref{bosonmass2}) and (\ref{condensate1}), one finds
\beeq
\mu_1^2 - \mu^2 = - 4\pi m \la \psibar\psi \ra_\theta ~~,
\label{PCAC}
\eneq
which is a PCAC relation.  

The Polyakov loop is, from (\ref{PWcorrespondence}), 
\beqn
\la P_q \ra_{\theta,T} &=& \la e^{i(q/e) \wil} \ra_{\theta,L=T^{-1}}\cr
\noalign{\kern 8pt}
&=& \cases{0 &for $\myfrac{q}{e} \not=$ an integer\cr
\mybig \int_{-\infty}^\infty dp_W \, f(p_W)^* f(p_W - \myfrac{q}{e})
&for $\myfrac{q}{e} =$ an integer.\cr}
\label{Polyakov1}
\eeqn

The potential, $V(p_W)$, in Eq.\ (\ref{Schro2}) consists of two terms;
harmonic oscillator and cosine potentials.  The strength of the cosine 
term, $\kappa$, is
\beeq
\kappa = \cases{
e^\gamma m \mu_1 L^2 &for $\mu_1 L\gg 1$\cr
4\pi mL&for $\mu_1 L\ll 1$.\cr} 
\label{kappa1}
\eneq
Depending on relative
strength, the behavior of the ground state wave function is quite 
different.   If $\omega^2 \gg \kappa$, the potential
is approximated by the harmonic term, ie.\ 
$f(p_W)= (\omega/\pi)^{1/4} e^{-\omega p_W^2 /2}$.   The condition is 
satisfied   if  $m/\mu \ll 1 \ll \mu L$ or 
if  $m/\mu \ll \mu L \ll 1$.  In this regime
$\la \cos(\theta-2\pi p_W) \ra_f = e^{-\pi/\mu L} \cos \theta$ so that
\beqn
&&\mu_1 = \sqrt{\mu^2+(me^\gamma \cos\theta)^2} + me^\gamma
\cos\theta\cr 
\noalign{\kern 6pt}
&&\la \psibar\psi\ra_\theta = - {e^\gamma\over 2\pi}\, \mu_1
\cos\theta \cr
\noalign{\kern 6pt}
&&\la P_e\ra_\theta = e^{-\pi \mu /4T} \sim 0 \hskip 2cm
\for {m\over \mu} \ll 1 \ll \mu L= {\mu\over T}
\label{limit1}
\eeqn
and
\beqn
&&\mu_1^2 = \mu^2 + {8\pi m\over L} \, e^{-\pi/\mu L} \cos\theta \cr
\noalign{\kern 6pt}
&&\la \psibar\psi\ra_\theta = - {2\over L} \, e^{-\pi/\mu L} \,
\cos\theta \cr
\noalign{\kern 6pt}
&&\la P_e\ra_\theta = e^{-\pi \mu /4T} \sim 1  \hskip 2cm
\for {m\over \mu} \ll  \mu L = {\mu\over T} \ll 1 ~~,
\label{limit2}
\eeqn

In the opposite limit $\kappa \gg \omega^2$, the cosine
term dominates.  However, the harmonic potential cannot be ignored
as it lifts the degeneracy of  the cosine potential.
In the large  volume  limit 
 the harmonic potential selects one of the minima of
the cosine potential at $p_W=\bar\theta/2\pi$ where $\bar\theta= \theta
-2\pi[(\theta+\pi)/2\pi]$.  Hence  $V \sim  \onehalf  e^\gamma m
\mu_1 L^2 ( 2\pi p_W - \bar\theta)^2$ so that
$f(p_W) = (\tilde\mu L)^{1/4} e^{-\tilde\mu L(2\pi
p_W-\bar\theta)^2/8\pi}$ where $\tilde\mu^2 = 2m\mu_1 e^\gamma$.
[For $\theta\sim \pi ~ (mod~2\pi)$, $f$ has two peaks
at $p_W \sim \pm \onehalf$.]  This leads to $\la \cos(\theta-2\pi p_W)\ra_f
 = e^{-\pi/\tilde\mu L} \sim 1$.   Consequently
\beqn
&&\mu_1 = \tilde\mu = 2 e^\gamma m \cr
\noalign{\kern 6pt}
&&\la \psibar\psi \ra_\theta = - {e^{2\gamma}\over \pi} \, m \cr
\noalign{\kern 6pt}
&&\la P_e\ra_\theta = e^{-\pi e^\gamma m  / 2T} 
\hskip 1cm \for m \gg \mu ~~,~~ \mu L = {\mu\over T} \gg 1 ~.
\label{limit3}
\eeqn
Notice that the chiral condensate increases linearly with $m$.
However,  there is no
$\theta$ dependence to the leading order.  (See the previous footnote, too.)

At high temperatures ($T=L^{-1} \gg \mu$), $\omega^2/\kappa= 
\pi \mu^2/4mT$.  So long as $m\not= 0$, eventually the cosine term
dominates.  However, the both terms in the potential become small
in this limit.  A good estimate is obtained by treating the cosine
term as a perturbation.  Numerical evaluation supports the result.
(See fig.\ 3 below.)

Write Eq.\ (\ref{Schro2}) as $(H_0 + V') f = \ep f$ where 
$V'$ is the cosine term.  Eigenstates of $H_0$ are denoted by 
$\{ |n \ra \}$ ($n=0,1,2,\cdots$).   Then
\beeq
\la n | \cos{(\theta - 2\pi p_W  )} | 0 \ra = 
{{(e^{i\theta } + {(-1)}^{n} \, e^{-i\theta })}\over2 } \, e^{-\pi / \mu L} \, 
{{1}\over{\sqrt{ n!} }} \, 
{\Big( {{2\pi }\over{\mu L}}  \Big) }^{{n}\over{2}}~. 
\eneq
$f$ is given by
$|f \ra = |0\ra - \sum_{n=1}  |n\ra \la n | V'|0 \ra / 2\omega n$
so that
\beqn
\la \cos(\theta- 2\pi p_W) \ra_f = e^{-\pi^2/\omega} \, \cos\theta 
    \hskip 5.5cm &&\cr
\noalign{\kern 14pt}
 + {\kappa\over 2\omega} \, e^{-2\pi^2/\omega}
\Bigg\{ \int_0^{2\pi^2/\omega} dz \, {e^z-1\over z}
+ \cos 2\theta \int_0^{2\pi^2/\omega} dz \, {e^{-z} -1\over z} \Bigg\}~. &&
\label{cosine1}
\eeqn

For $T=L^{-1} \ll \mu$ ($\omega \gg 1$) the first term in (\ref{cosine1})
dominates over the rest to reproduce (\ref{limit1}).  For $T \gg \mu$
($\omega \ll 1$),  $\int_0^{2\pi^2/\omega} dz \, (e^z-1)/z \sim
(\omega/2\pi^2) e^{2\pi^2/\omega}$ so that 
$\la \cos(\theta- 2\pi p_W) \ra_f \sim m/\pi T$.   We obtain
\beqn
&&\mu_1^2 = \mu^2 + 8m^2 \cr
\noalign{\kern 6pt}
&&\la \psibar\psi\ra_\theta = - {2m\over \pi} \cr
\noalign{\kern 6pt}
&&\la P_e\ra_\theta = e^{-\pi \mu /4 T} \bigg( 1- {6.60\,  m\over \mu} \,
e^{-\pi T /\mu} \, \cos{\theta } \bigg)  
\hskip .5cm \for T \gg \mu, m.
\label{limit4}
\eeqn
\begin{figure}[t]
\hskip 2cm
\epsfxsize= 10.cm
\epsffile[79 414 524 713]{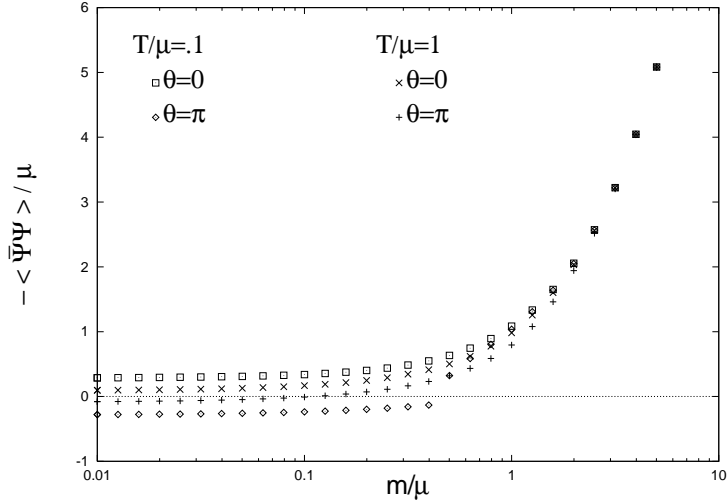}
\vskip -0cm
\caption{$m/\mu$ dependence of chiral condensates at given $T/\mu$ 
and $\theta$.}
\label{fig.2}
\vskip 0.3cm
\end{figure}

In the intermediate range of parameter values Eqs.\ (\ref{Schro2})
and (\ref{bosonmass2}) must be solved numerically.  The computational
algorithm is the following.  With given $(\mu L, m/\mu, \theta)$, we
first assign an input value for $\kappa=\kappa_{\rm in}$.   The potential
in (\ref{Schro2}) is specified with ($\mu L, \kappa_{\rm in}, \theta$).
Eq.\ (\ref{Schro2}) determines $f(p_W)$, from which one can determine
$\mu_1$, solving (\ref{bosonmass2}).   With this new $\mu_1$ one 
recomputes $\kappa=\kappa_{\rm out}$ by (\ref{Schro3}).  Schematically
\beeq
\kappa_{\rm in} \go V(p_W) \go f(p_W) \go \mu_1 \go \kappa_{\rm out}~.
\label{mapping1}
\eneq
$\kappa_{\rm out}$ must coinsides $\kappa_{\rm in}$. This  gives
a consistency condition to determine $\kappa$ with given 
$(\mu L, m/\mu, \theta)$.

In fig.\ 2 chiral condensates are plotted as functions of
$m/\mu$.   Asymptotically ($m\gg \mu$) the behavior is given by
(\ref{limit3}).  
$T$ dependence of chiral condensates is displayed in fig.\ 3. \cite{RHHI}
  Low and
high temperature limits agree with (\ref{limit1}) and (\ref{limit4}).
The $T$ dependence of chiral 
condensates with given $m$ and $\theta$ is smooth. 
This is consistent with the Mermin-Wagner theorem that in a one-dimensional
system there is no phase transition at finite temperature. \cite{Mermin}  

\begin{figure}[tb]
\hskip 2cm
\epsfxsize= 10.cm
\epsffile[79 410 526 721]{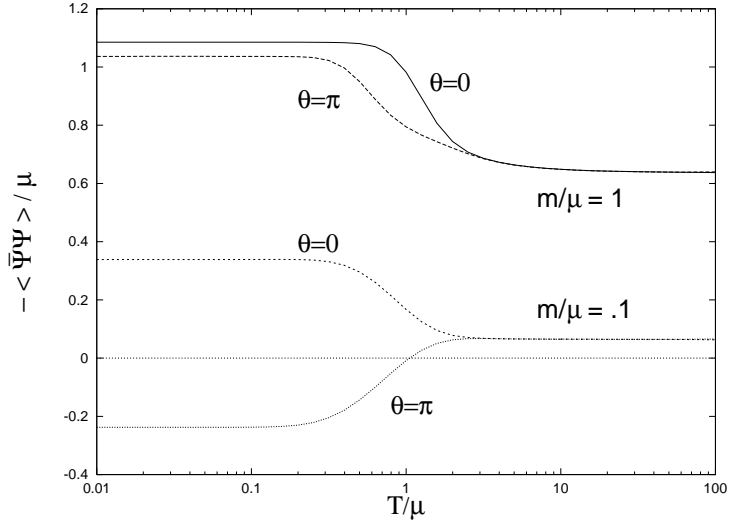}
\vskip -0cm
\caption{$T$ dependence of chiral condensates at fixed $m/\mu$ and
$\theta$.}
\label{fig.3}
\vskip 0.3cm
\end{figure}

$T$ dependence of Polyakov loops is displayed with various
values of $m/\mu$ in fig.\ 4.   Notice that at $\theta=0$ the curve
smoothly changes as $m/\mu$.  However, around $\theta=\pi$ nontrivial $T$
dependence is  observed for $m \ge \mu$.  The origin of this behavior is
traced back to the wave function $f(p_W)$ in the corresponding problem
on $S^1$ with $L=T^{-1}$.   $f(p_W)$ at moderately low temperature
has two dominant peaks at $p_W=\pm(1-\ep)$ around $\theta=\pi$,
whereas it has only one dominant peak at $p_W=0$ for $\theta=0$.   As
(\ref{Polyakov1}) shows, the Polyakov loop is determined by the overlap
of $f(p_W)^*$ and $f(p_W -1)$.  Hence it vanishes quickly for $\theta=0$,
but approaches $\sim .5$ for $\theta\sim\pi$ at low, but moderate $T$.
As $T$ further gets lowered, the two peaks become narrower and sharper.
When the width of the peaks becomes smaller than $\ep$, the overlap of
the wave functions and  Polyakov loop vanish.  If $\theta$ is not exactly
$\pi$, but is very close to $\pi$,   the asymmetry in the potential,
enhanced by the factor $m/T^2$, 
becomes important at sufficiently low $T$ and the wave function has a sharp
peak around one of the minima.  See the wave functions displayed in fig.\ 1.
In the numerical evaluation presented in fig.\ 4, the computer picks a value
for $\theta$ which is not exactly $\pi$.  The transition from the plateau
($\sim .5$) to zero at $\theta \sim \pi$ is caused by this change of the wave
function.   This explains the two-step behavior observed in fig.\ 4. 

\begin{figure}[bt]
\vskip .3cm
\hskip 2cm
\epsfxsize= 10.cm
\epsffile[96 414 524 713]{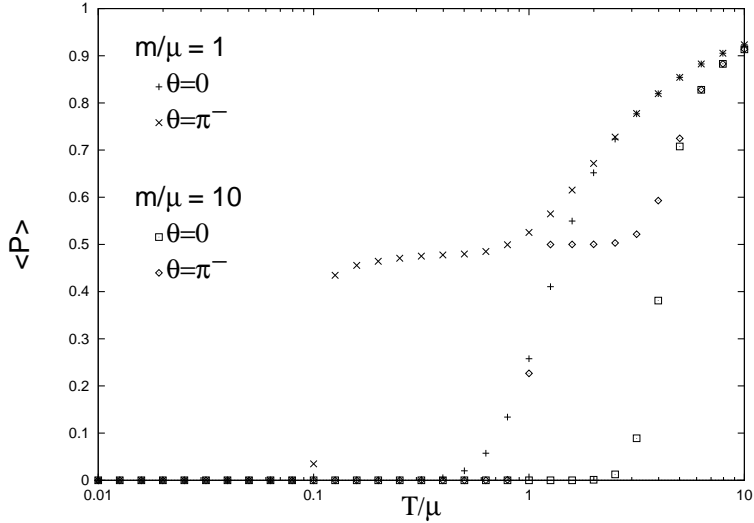}
\vskip -0cm
\caption{$T/\mu$ dependence of Polyakov loops with given $m/\mu$ and
$\theta$.   Two-step behavior is observed at $\theta \sim \pi$.}
\label{fig.4}
\end{figure}

In fig.\ 3 we observe that the pattern of the $T$ dependence of
$\la\psibar\psi\ra$ changes as $m/\mu$ is increased.  At $\theta=\pi$,
$\la\psibar\psi\ra/\mu$ decreases (increases) as $T/\mu$ increases 
for $m/\mu=.1$ ($m/\mu=1$).

Indeed there arises a discontinuity in the $m$ dependence of chiral
condensates at low temperature at $\theta=\pi$, at least in our
approximation scheme.   We have displayed it  at $T/\mu=0.03$ in fig.\ 5.  
We observe that $\la\psibar\psi\ra/\mu$ discontinuosly changes at $m/\mu=.437$.

\begin{figure}[bt]
\vskip .3cm
\hskip 2cm
\epsfxsize= 9.cm
\epsffile[82 411 527 722]{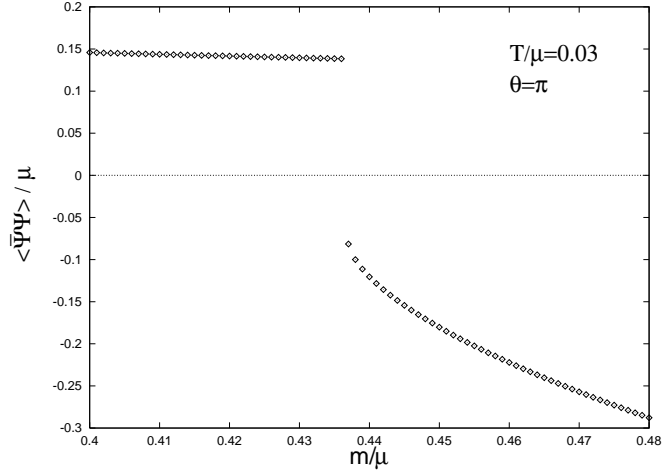}
\vskip -0cm
\caption{A discontinuity in  chiral
condensates is observed at the $m/\mu \sim .437$ when $T/\mu=0.03$ and
$\theta=\pi$.}
\label{fig.5}
\end{figure}

To understand the origin of the discontinuity, we consider a solution
$\kappa_{\rm in} =\kappa_{\rm out}$ in (\ref{mapping1}).  Write $\kappa_{\rm
out} = g(\kappa_{\rm in})$.  We are looking for a solution to
$g(\kappa) =\kappa$, or a fixed point $\kappa_{\rm fix}$ of $g(\kappa)$.  
As $m/\mu$ varies with given $T/\mu$ and $\theta$, $g(\kappa)$, and therefore
$\kappa_{\rm fix}$ change. At $\theta=\pi$, there is a critical value for
$m/\mu$ at which the fixed point bifurcates.  In a certain range of $m/\mu$,
there appear three fixed points, two stable and one unstable.  Among the two
stable fixed points, one of them has a larger chiral condensate and 
therefore a lower energy density, corresponding to the vacuum.
This bifurcation induces a discontinuous change in chiral condensates
as $m/\mu$ varies.  We have displayed the mapping 
$\kappa_{\rm out}=g(\kappa_{\rm in})$ in the critical
region in fig.\ 6 (a).  We observe that saddle node bifurcation takes place at
$m/\mu=0.4368$.

At very low $T$ the critical mass $m_c/\mu$ can be determined analytically. 
Suppose that $m={\rm O}(\mu)$.  For $\mu L= \mu/T \gg 1$, 
$\kappa=e^\gamma m\mu_1 L^2$.  At $\theta=\pi$, 
\beeq
V = (\pi\mu L p_W)^2 + \kappa \cos 2\pi p_W ~~. 
\label{critical1}
\eneq
There is always a solution which satisfies 
$(\mu L)^2 > 2\kappa$ or $\mu^2 > 2 e^\gamma m \mu_1$.  In this case
 $f(p_W)$ is sharply localized around $p_W=0$.  This yields
\beeq
\mu_1 = \sqrt{ \mu^2 +(me^\gamma)^2 } - me^\gamma ~~.
\label{critical2}
\eneq

If $(\mu L)^2 < 2\kappa$ or $\mu^2 < 2 e^\gamma m \mu_1$, 
 $f(p_W)$ is localized around $\pm \bar p_W\not= 0$ where
\beqn
&&2\pi\bar p_W = {2m\mu_1 e^\gamma\over \mu^2} \, \sin 2\pi\bar p_W \cr
&&\mu_1 = \sqrt{ \mu^2 + (me^\gamma \cos 2\pi \bar p_W )^2 }
- me^\gamma \cos 2\pi \bar p_W 
\label{critical3}
\eeqn
A solution to (\ref{critical3}) exists only for $ m/\mu > 0.435 $. 
This solution corresponds to a bigger $\mu_1$ or $-\la \psibar\psi\ra$,
and therefore to a lower energy density.  In other words
$m_c/\mu= 0.435$ at $T=0$.  Numerically we have found $m_c/\mu=0.435, 0.437$, 
and 
0.454 at  $T/\mu=0.01, 0.03$ and 0.07, respectively.  Above $T/\mu=0.12$ the 
function $g(x)$ has only one fixed point for all values of $m/\mu$ so that the
discontinuity disappears.

\begin{figure}[tb]
\hskip 1.cm
\epsfxsize= 13.cm
\epsffile[76 507 544 723]{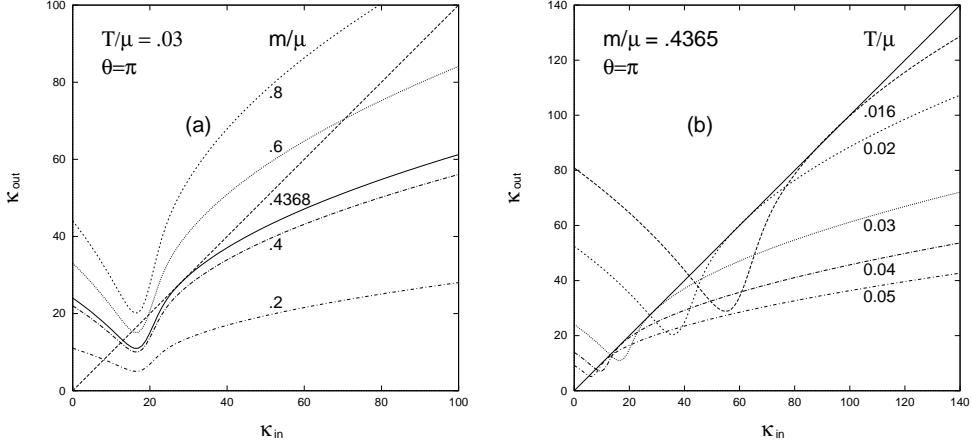}
\vskip -0cm
\caption{$\kappa_{\rm in}$-$\kappa_{\rm out}$ plots.  
(a)  At $T/\mu=0.03$ and $\theta=\pi$.  As $m/\mu$ changes, the 
     fixed point bifurcates at $m/\mu=0.4368$.
(b)  With a fixed $m/\mu=0.4365$, $T/\mu$ is varied.  The curves
     remain almost critical.}
\label{fig.6}
\vskip 0.3cm
\end{figure}

It is not clear, however, if the discontinuity discovered above is real in
the full theory, or just an artifact of the approximation in use.  
In determing the boson mass $\mu_1$ in (\ref{bosonmass1}) and
(\ref{bosonmass2}) we have ignored nonlinear terms in the $\phi$ field
in $H_\mass$, retaining only the $\phi^2$ term.  Those nonlinear terms
are expected to affect the boson mass.  

Furthermore, the Mermin-Wagner theorem ensures that there should be no
discontinuity in $\la\psibar\psi\ra$ when $T/\mu$ varies with $m/\mu$ kept
fixed.   The discontinuity in the $m/\mu$ dependence is consistent with
the Mermin-Wagner theorem only if $m_c/\mu$ is universal, being independent of
$T/\mu$.  Numerically we have found that $m_c/\mu$ is almost universal,
although there appears tiny dependence on $T/\mu$.   In fig.\ 6 (b) we have
plotted $\kappa_{\rm in}$-$\kappa_{\rm out}$ with a fixed $m/\mu=.4365$ at
various $T/\mu$.  The curve is critical at $T/\mu\sim 0.03$, while it is
slightly off at higher or lower temperature despite it may be very hard
to see visually.  It is striking that the critical value $m_c/\mu$ is
almost insensitive to $T/\mu$ to such a degree of accuracy.

There are two possible senarios when the whole interactions are taken into
account.  It may turn out that the discontinuity in $m/\mu$ disappears, 
being replaced by a crossover transition.  Or the discontinuity is real,
taking place at a universal mass value $m_c/\mu$. Further investigation is
necessary to
determine which picture is right.

In this paper we have evaluated $T$- and $m$-dependence of chiral condensates
and Polyakov loops.  We have demonstrated that at $\theta$$\sim$$\pi$
there appears anomalous behavior when $m/\mu$ is 0.4 $\sim$ 0.5.  
Mathematically these anomalous phenomena are
related to the bifurcation process in the solutions of the vacuum equation.
Electromagnetic interactions and fermion mass collaborate to induce 
anomalous behavior of the vacuum solutions.   This is reminessent
of chaotic dynamics in nonlinear systems.    More detailed analysis will be
reported separately.

\bigskip

\leftline{\bf Acknowledgements}
This work was supported in part  by the U.S.\ Department of Energy
under contracts DE-AC02-83ER-40105 (Y.H.) and
DE-FG02-87ER-40328 (R.R.).  One of the authors (Y.H.) would like to thank 
Physics Department, Brookhaven National Laboratory for its hospitality
where a part of the work was carried out.


\def\ap {{\it Ann.\ Phys.\ (N.Y.)} }
\def\cmp {{\it Comm.\ Math.\ Phys.} } 
\def\ijmpA {{\it Int.\ J.\ Mod.\ Phys.} {\bf A}} 
\def\ijmpB {{\it Int.\ J.\ Mod.\ Phys.} {\bf B}} 
\def\ijmpC {{\it Int.\ J.\ Mod.\ Phys.} {\bf C}} 
\def\jmp {{\it  J.\ Math.\ Phys.} } 
\def\mplA {{\it Mod.\ Phys.\ Lett.} {\bf A}} 
\def\mplB {{\it Mod.\ Phys.\ Lett.} {\bf B}} 
\def\plB {{\it Phys.\ Lett.} {\bf B}} 
\def\plA {{\it Phys.\ Lett.} {\bf A}} 
\def\nc {{\it Nuovo Cimento} } 
\def\npB {{\it Nucl.\ Phys.} {\bf B}} 
\def\pr {{\it Phys.\ Rev.} } 
\def\prl {{\it Phys.\ Rev.\ Lett.} } 
\def\prB {{\it Phys.\ Rev.} {\bf B}} 
\def\prD {{\it Phys.\ Rev.} {\bf D}} 
\def\prp {{\it Phys.\ Report} } 
\def\ptp {{\it Prog.\ Theoret.\ Phys.} } 
\def\rmp {{\it Rev.\ Mod.\ Phys.} } 

\vskip .7cm
\leftline{\bf References}

\renewenvironment{thebibliography}[1]
	{\begin{list}{[$\,$\arabic{enumi}$\,$]}  
	{\usecounter{enumi}\setlength{\parsep}{0pt}
	 \setlength{\itemsep}{0pt}  \renewcommand{\baselinestretch}{1.2}
         \settowidth
	{\labelwidth}{#1 ~ ~}\sloppy}}{\end{list}}

\myend
\begin{thebibliography}{99}


\small

\bibitem{Schwinger}
 J. Schwinger, \pr {\bf 125} (1962) 397 ;  {\bf 128} (1962) 2425.
\bibitem{Lowenstein} 
J.H. Lowenstein and J.A. Swieca, \ap {\bf 68} (1971) 172.
\bibitem{Casher} 
A. Casher, J. Kogut and L. Susskind, \prl {\bf 31} (1973) 792; 
\prD {\bf 10} (1974) 732 .
\bibitem{CJS}
S. Coleman, R. Jackiw, and L. Susskind,  \ap {\bf 93} (1975) 267.
\bibitem{Coleman2} S.\ Coleman, \prD {\bf 11} (1975) 2088.
\bibitem{Coleman1} S.\ Coleman,  \ap {\bf 101} (1976) 239.
\bibitem{Nielsen} N.K. Nielsen and B. Schroer, \npB {\bf 120} (1977) 62.


\bibitem{Wolf}  D. Wolf and J. Zittartz, {\it Z. Phys.} {\bf B59} (1985) 117.
\bibitem{Manton}  N. Manton, \ap {\bf 159} (1985) 220.

\bibitem{HH} J.E. Hetrick and Y. Hosotani, \prD {\bf 38} (1988) 2621.
\bibitem{Iso} S. Iso and H. Murayama, \ptp {\bf 84} (1990) 142.

\bibitem{Link} R. Link, \prD {\bf 42} (1990) 2103.
\bibitem{Sachs} I.\ Sachs and A. Wipf, {\it Helv. Phys. Acta.} 
{\bf 65} (1992) 652 .

\bibitem{Joos} H. Joos, {\it Helv. Phys. Acta.} {\bf 63} (1990) 670,
 H. Joos and S.I. Azakov, {\it Helv. Phys. Acta.} {\bf 67} (1994) 723.

\bibitem{Ellis} J.\ Ellis, Y.\ Frishman, A.\ Hanany, M.\ Karliner, 
       \npB {\bf 382} (1992) 189.  
\bibitem{Ross} M.B. Paranjape and R. Ross, \prD {\bf 48} (1993) 3891.
\bibitem{Paranjape} M.B. Paranjape,  \prD {\bf 48} (1993) 4946.

\bibitem{HHI}  J.E.\ Hetrick, Y.\ Hosotani and S.\ Iso, \plB {\bf 350} 
(1995) 92;  \prD {\bf 53} (1996) 7255;
Y.\ Hosotani, {\tt hep-ph/9510387}. 

\bibitem{HNZ}  T.H. Hansson, H.B. Nielsen and I. Zahed,
        \npB {\bf 451} (1995) 162.
\bibitem{Gross} D.J.\ Gross, I.R.\ Klebanov, A.V.\ Matytsin, A.V.\ Smilga, 
        \npB {\bf 461} (1996) 109.
\bibitem{Grignani1} G.\ Grignani, G.\ Semenoff, P.\ Sodano, O.\ Tirkkonen, 
      {\tt hep-th/9511110}.


\bibitem{Ramon}  R.\ Rodriguez and Y.\ Hosotani, \plB {\bf 375} (1996)
273.
\bibitem{HRHI}  Y.\ Hosotani, R.\ Rodriguez, J.E.\ Hetrick and S.\
Iso, {\tt hep-th/9606129}.

\bibitem{RHHI}  R.\ Rodriguez, Y.\ Hosotani, J.E.\ Hetrick and S.\
Iso, {\tt hep-th/9608123}.





\bibitem{Irving} A.C. Irving and J.C. Sexton, in {\it Lattice 1995},
 \npB {\bf 47} (Proc.\ Suppl.) (1996) 679;
I.\ Horvath,   \npB {\bf 47} (Proc.\ Suppl.) (1996) 683;
V.\ Azcoiti et al., \npB {\bf 47} (Proc.\ Suppl.) (1996) 687.

\bibitem{Horvath} I. Horvath, \prD {\bf 53} (1996) 3808. 

\bibitem{Harada} K. Harada, T. Sugiura and M. Taniguchi,
  \prD {\bf 49} (1994) 4226;  
K.\ Harada, A.\ Okazaki, and M.\ Taniguchi,  {\tt hep-th/9509136}.

\bibitem{Mermin} N.D.\ Mermin and H.\ Wagner, \prl {\bf 17} (1966)
1133.

\end{thebibliography}
